\documentclass[%
 reprint,
 amsmath,amssymb,
 aps,pra,
 superscriptaddress
]{revtex4-2}
\usepackage{graphicx}
\usepackage{amsmath}
\usepackage{siunitx}
\usepackage{nicefrac}
\usepackage[hidelinks]{hyperref}

\usepackage{lipsum}

\begin{document}
\title{Interactions enable Thouless pumping in a nonsliding lattice
}
\author{Konrad Viebahn}
\thanks{These authors contributed equally. Correspondence should be addressed to K.V. (viebahnk@phys.ethz.ch).}
\affiliation{Institute for Quantum Electronics \& Quantum Center, ETH Zurich, 8093 Zurich, Switzerland
}
\author{Anne-Sophie Walter}
\thanks{These authors contributed equally. Correspondence should be addressed to K.V. (viebahnk@phys.ethz.ch).}
\affiliation{Institute for Quantum Electronics \& Quantum Center, ETH Zurich, 8093 Zurich, Switzerland
}
\author{Eric Bertok}
\thanks{These authors contributed equally. Correspondence should be addressed to K.V. (viebahnk@phys.ethz.ch).}
\affiliation{Institute for Theoretical Physics, Georg-August-Universität Göttingen, 37077 Göttingen, Germany}
\author{Zijie Zhu}
\affiliation{Institute for Quantum Electronics \& Quantum Center, ETH Zurich, 8093 Zurich, Switzerland
}
\author{Marius Gächter}
\affiliation{Institute for Quantum Electronics \& Quantum Center, ETH Zurich, 8093 Zurich, Switzerland
}
\author{Armando A. Aligia}
\affiliation{Instituto de Nanociencia y Nanotecnología CNEA-CONICET, Centro Atómico Bariloche and Instituto Balseiro, 8400 Bariloche, Argentina}
\author{Fabian Heidrich-Meisner}
\affiliation{Institute for Theoretical Physics, Georg-August-Universität Göttingen, 37077 Göttingen, Germany}
\author{Tilman Esslinger}
\affiliation{Institute for Quantum Electronics \& Quantum Center, ETH Zurich, 8093 Zurich, Switzerland
}

\begin{abstract}
A topological `Thouless' pump represents the quantised motion of particles in response to a slow, cyclic modulation of external control parameters.
The Thouless pump, like the quantum Hall effect, is of fundamental interest in physics because it links physically measurable quantities, such as particle currents, to geometric properties of the experimental system, which can be robust against perturbations and thus technologically useful.
So far, experiments probing the interplay between topology and inter-particle interactions have remained relatively scarce.
Here we observe a Thouless-type charge pump in which the particle current and its directionality inherently rely on the presence of strong interactions.
Experimentally, we utilise a two-component Fermi gas in a dynamical superlattice which does not exhibit a sliding motion and remains trivial in the single-particle regime.
However, when tuning interparticle interactions from zero to positive values, the system undergoes a transition from being stationary to drifting in one direction, consistent with quantised pumping in the first cycle.
Remarkably, the topology of the interacting pump trajectory cannot be adiabatically connected to a non-interacting limit, highlighted by the fact that only one atom is transferred per cycle.
Our experiments suggest that Thouless charge pumps are promising platforms to gain insights into interaction-driven topological transitions and topological quantum matter.
\end{abstract}

\maketitle

\section{Introduction}

An adiabatic change of an energy landscape represents one of the simplest strategies to induce controlled particle motion.
For example, a sliding potential minimum can carry a trapped particle from one point to another, both in classical and quantum mechanics.
However, the wave nature of quantum-mechanical states allows for additional physical effects arising from a geometric phase change when the Hamiltonian is time-dependent~\cite{xiao_berry_2010}.
A geometric or `Berry' phase is not usually evident from the underlying potential but requires knowledge of eigenstates and their geometric structure.
Thouless showed that in a periodic system the geometric phase contributions after one adiabatic cycle sum to integers which correspond to the singularities enclosed by the pump trajectory~\cite{thouless_quantization_1983}.
Physically, these integers describe the quantisation of transported charge.
This phenomenon, known as Thouless charge pumping, is topologically protected against perturbations that are small compared to the energy gap between ground and excited states~\cite{xiao_berry_2010,asboth_short_2016,cooper_topological_2019}.
Therefore, topological charge pumps may also become technologically relevant as sources of quantised current~\cite{pekola_single-electron_2013}.
An alternative (but equivalent) description of Thouless pumping regards the time-periodic variation as a Floquet drive~\cite{oka_floquet_2019} which gives rise to a synthetic dimension of photon states.
In this two-dimensional picture, the topological pump represents a quantised bulk Hall drift.

While many aspects of topological band structures were realised with optical lattices with engineered gauge potentials~\cite{jotzu_experimental_2014,aidelsburger_measuring_2015,tarnowski_measuring_2019}, accessing the interacting regime still poses a significant challenge.
The reason is the unavoidable heating in a driven many-body system~\cite{reitter_interaction_2017,viebahn_suppressing_2021} and the problem of loading the bulk.
Therefore, only a few experiments have explored interacting topological quantum states in optical lattices~\cite{zhou_observation_2023,leonard_realization_2023,walter_quantization_2023}.
Experiments with Thouless pumps circumvent these experimental issues and, as we will demonstrate here, enable the study of interaction effects on the topology of many-body systems.
Theoretical papers have suggested a variety of interaction-related effects in pumping~\cite{niu_quantised_1984,berg_quantized_2011,qian_quantum_2011,wang_topological_2013,grusdt_realization_2014,zeng_fractional_2016,tangpanitanon_topological_2016,li_finite-size_2017,ke_multiparticle_2017,taddia_topological_2017,lindner_universal_2017,nakagawa_breakdown_2018,stenzel_quantum_2019,haug_topological_2019, unanyan_finite-temperature_2020, greschner_topological_2020, chen_simulating_2020,fu_nonlinear_2022,mostaan_quantized_2022,gawatz_prethermalization_2022,andrews_stability_2021}.
So far, interactions did not play a major role in pumping experiments~\cite{citro_thouless_2023,kraus_topological_2012,lohse_thouless_2016,nakajima_topological_2016,ke_topological_2016,lu_geometrical_2016,ma_experimental_2018,cerjan_thouless_2020,nakajima_competition_2021,minguzzi_topological_2022,xiang_simulating_2023}.
Recently, the effects of interactions on pumping have been explored in two experimental platforms, that is, in the mean-field regime~\cite{jurgensen_quantized_2021,jurgensen_quantized_2023} and in a Fermi-Hubbard system~\cite{walter_quantization_2023}.
In both cases, interactions which exceeded the protecting energy gap caused a suppression of the quantised pumping motion. 
However, the question of whether interactions can cause or even stabilise topological behaviour has largely remained unanswered on the experimental level~\cite{dzero_topological_2016,lin_interaction-induced_2020,kuno_interaction-induced_2020,bertok_splitting_2022,ostrovsky_interaction-induced_2010,wang_interaction-induced_2012,budich_fluctuation-driven_2013,kumar_interaction-induced_2016,salerno_interaction-induced_2020,zheng_measuring_2020,herbrych_interaction-induced_2021,luntama_interaction-induced_2021}.

Here, we report on the experimental observation of interaction-induced charge pumping using interacting fermionic atoms in a dynamically modulated optical lattice.
The optical lattice realises a Hubbard model with modulated hopping matrix elements and onsite potentials. 
For Hubbard interactions larger than a nonzero critical value, the atoms exhibit a pumping motion, while they remain stationary in the non-interacting limit.
Our measurements are consistent with a quantised displacement of the atomic cloud in the first pump cycle, in quantitative agreement with time-dependent matrix-product-state (MPS) simulations~\cite{schollwock_density-matrix_2005,schollwock_density-matrix_2011}.
Interestingly, the transferred charge per pump cycle is half of its usual value in a non-interacting system, as predicted in Ref.~\cite{bertok_splitting_2022}, and the pumping mechanism does not have a classical counterpart. At very large interactions and beyond a second critical interaction strength, the pumped charge vanishes again. Crucially,
the region with pumping of one charge per cycle cannot be adiabatically connected to the non-interacting limit.
Our work establishes an example of topological phase transitions controlled by the interaction strength.
Previously, topological transitions were observed by tuning external parameters in non-interacting models, such as those which break inversion-symmetry in the Haldane model~\cite{jotzu_experimental_2014,tarnowski_measuring_2019,wintersperger_realization_2020}.

The pump involves the physics of the ionic Hubbard model~\cite{fabrizio_band_1999,torio_phase_2001,manmana_quantum_2004,torio_quantum_2006,pertot_relaxation_2014,messer_exploring_2015}, which gives rise to the observed transitions.  Due to the inherent SU(2) symmetry in the spin sector, this model possesses a Mott-insulating region with gapless spin excitations.
Since the pump realised in our experiment traverses through this region, strictly  speaking, there cannot be adiabatic pumping and quantised pumping must eventually break down~\cite{nakagawa_breakdown_2018}.
Notably, though, the experimental data, as well as our numerical simulations for realistic conditions show that the transferred charge in the first pump cycle remains robust and quantised.  
Moreover, an analysis of correlations illustrates that  initial spin excitations are converted into defects in the charge sector with a time delay and therefore do not affect the pumped charge immediately, thus explaining the experimental observation. Therefore, a Thouless pump in a two-component Rice-Mele model with Hubbard interactions allows to systematically study the coupling between the spin and charge sectors on experimentally and numerically accessible time scales.
Alternative approaches for interaction-induced pumping, including Ref.~\cite{lin_interaction-induced_2020}, involve only two particles and pumping of pairs, different from the many-body situation considered here where only one particle is pumped per cycle.

Our work goes substantially beyond the existing experiments on topology with interacting atoms in optical lattices \cite{leonard_realization_2023,walter_quantization_2023,zhou_observation_2023}.
Most importantly, the interaction-induced pumping is not adiabatically connected to the limit of vanishing Hubbard interactions, thus studying physics beyond previous Thouless pump experiments \cite{walter_quantization_2023}.

\section{Interaction-induced charge pump}

\begin{figure}[t!]
    \includegraphics[width=0.48\textwidth]{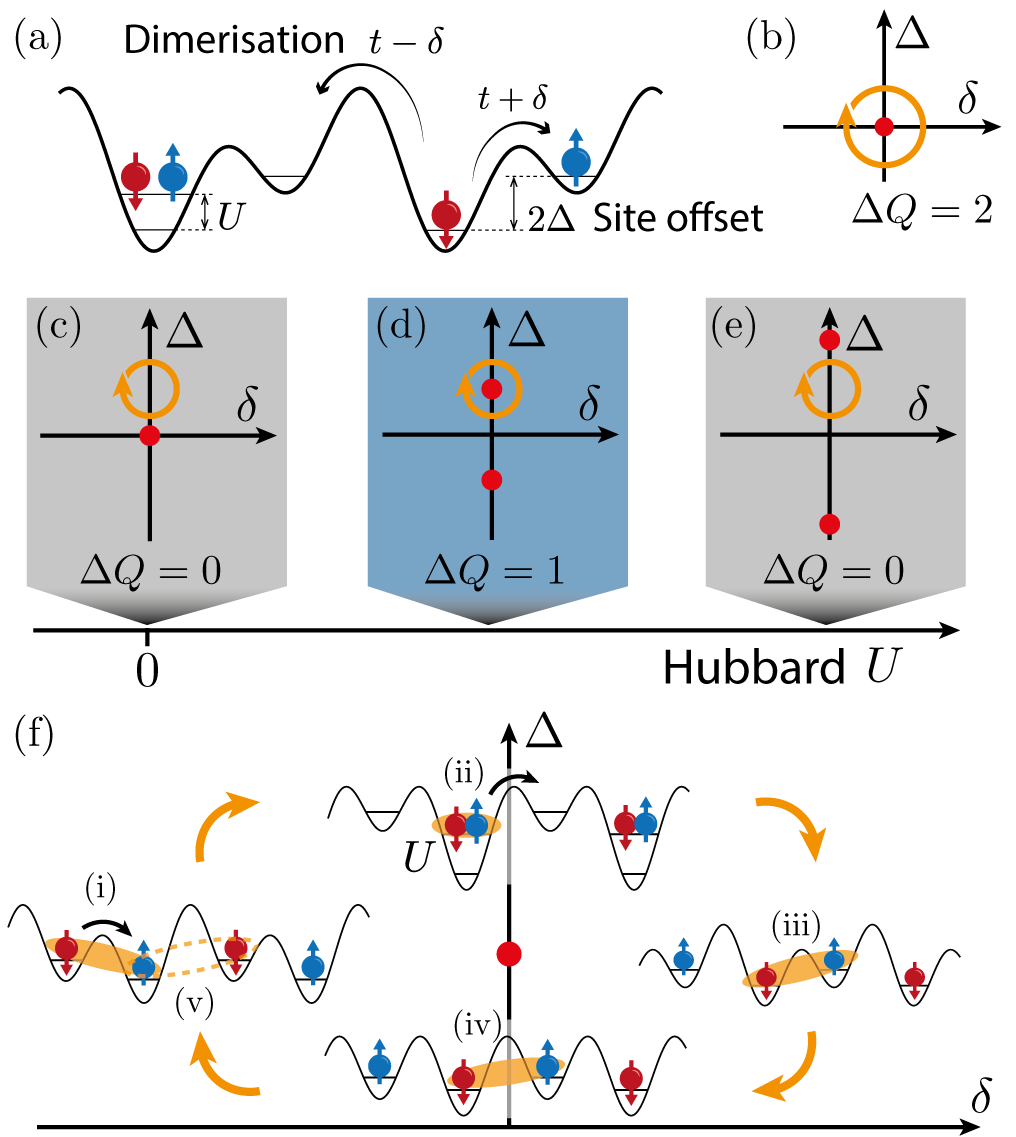}
         \caption{\textbf{Topological pumping induced by Hubbard interactions in a lattice without sliding motion.} (a) Dynamically modulated Rice-Mele-Hubbard model. The bond dimerisation ($\delta$) and the site offset ($\Delta$) are periodically modulated. (b) Pump trajectory centred at $\Delta = 0$, leading to a pumped charge of $\Delta Q = 2$ for spinful fermions at $U = 0$. (c-e) Pump trajectories centred at $\Delta > 0$. (c) For vanishing interactions ($U = 0$) the pumped charge is zero. (d) A finite Hubbard $U$ causes the splitting of singularities to $\Delta \gtrless 0$, which leads to interaction-induced topological pumping with $\Delta Q = 1$. (e) Once the Hubbard $U$ is too large, the pump is rendered topologically trivial.
         (f) Schematic illustration of the interaction-induced pump on the microscopic level. The red dot indicates the position of a single singularity at finite Hubbard $U$.}
    \label{fig:1}
\end{figure}

The specific Hamiltonian studied in this work is the Rice-Mele-Hubbard model
\begin{eqnarray}\label{eqn:RM}
    \hat{H}(\tau) &=& - \sum_{j,\sigma}\left[t + (-1)^j\delta(\tau)\right]\left(\hat{c}_{j\sigma}^\dagger \hat{c}_{j+1\sigma} + \text{h.c.}\right) \\
    \nonumber &&+\,\Delta(\tau)\sum_{j,\sigma} (-1)^j \hat{c}_{j\sigma}^\dagger \hat{c}_{j\sigma}+U\sum_{j} \hat{n}_{j\uparrow} \hat{n}_{j\downarrow}~,
\end{eqnarray}
which is parametrised by the bond dimerisation $\delta(\tau)$ and the sublattice site offset $\Delta(\tau)$, which both depend on time $\tau$ [Fig.~\ref{fig:1}(a)].
The fermionic annihilation and number operators for spin $\sigma \in \{\uparrow,\downarrow \}$ on lattice site $j$ are denoted by $\hat{c}_{j\sigma}$ and $\hat{n}_{j\sigma}$, respectively.
Let us first consider the non-interacting limit ($U = 0$) in which the whole parameter space is spanned by $\delta$ and $\Delta$ [Fig.~\ref{fig:1}(b)].
The relevant topological invariant for charge pumping is the (charge--) Berry phase of the lowest band which becomes singular at the origin of the $\delta$--$\Delta$ plane.
At half-filling, that is, two fermions per unit cell, a trajectory enclosing the singularity pumps a total of $\Delta Q = 2$ charges to the neighbouring unit cell per pump cycle (one spin-$\uparrow$, one spin-$\downarrow$).
For trajectories that do not enclose the singularity [Fig.~\ref{fig:1}(c)] the pumped charge is zero.

The Rice-Mele model at finite Hubbard $U$ gives rise to a rich many-body phase diagram at half-filling~\cite{fabrizio_band_1999,torio_phase_2001,manmana_quantum_2004,torio_quantum_2006,nakagawa_breakdown_2018,bertok_splitting_2022,manmana_topological_2012,aligia_topological_2023}.
The phases are governed by the competition and interplay of the parameters $U$, $\delta$, and $\Delta$. In short, the Su-Schrieffer-Heeger (SSH) lattice with $\Delta = 0$ leads to a dimerised Mott insulator for $U\gg t$ ~\cite{manmana_topological_2012}, whereas the ionic Hubbard model ($\delta = 0$) exhibits band insulating ($\Delta \gtrsim U/2$ for $U\gg t$) and Mott insulating ($\Delta \lesssim U/2)$ phases, with a small dimerised intermediate phase around $\Delta \sim U/2$~\cite{torio_phase_2001,manmana_quantum_2004,stenzel_quantum_2019}.
Recent numerical calculations predict a splitting of the non-interacting singularity at the origin $[(\delta,\Delta) = (0,0)]$ into two, for increasing values of Hubbard $U$~\cite{bertok_splitting_2022} (see also refs.~\cite{zhou_interaction_2017,yan_yang_2018}).
The new singularities each exhibit a $2\pi$ winding of the (charge--)Berry phase and should thus be amenable to topological charge pumping in the interacting regime.

\begin{figure*}[t!]
    \includegraphics[width=\textwidth]{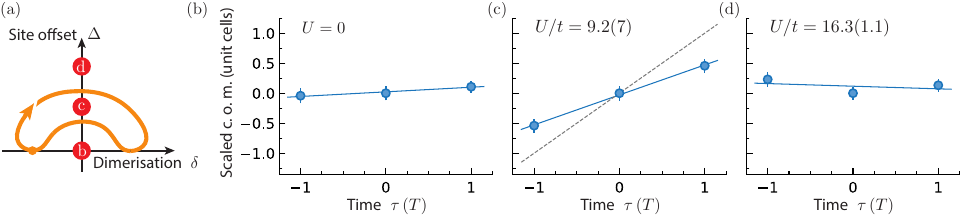}
         \caption{\textbf{Experimental observation of interaction-induced charge pumping.} Panel (a) shows a schematic representation of the experimental pump trajectory (see App.~\ref{app:exp} for details). The singularities are shown as red dots marked (b), (c), and (d), respectively, for increasing values of Hubbard $U$, corresponding to the following subplots. (b-d) The scaled centre-of-mass position (c.~o.~m.) is plotted as function of time $\tau$ for three different values of Hubbard $U$ (b): $U = 0$, (c): $U/t = 9.2(7)$, and (d): $U/t = 16.3(1.1)$.
         The dashed line in (c) represents a pumped charge of $\Delta Q = 2$, typical for the usual Rice-Mele pump~\cite{lohse_thouless_2016,nakajima_topological_2016,walter_quantization_2023}.
         Data points and error bars correspond to mean and standard error of 40 measurements.}
    \label{fig:2}
\end{figure*}

The key idea behind the interaction-induced pump considered here is the following: we choose a pump trajectory centred at a finite site offset $\Delta >0$, which does not enclose the singularity for $U = 0$.
This trajectory is topologically trivial in the non-interacting limit and transfers $\Delta Q = 0$ charges per pump cycle as sketched in Fig.~\ref{fig:1}(c).
Now, we successively tune the value of Hubbard $U$ until one of the two singularities (red dots in Fig.~\ref{fig:1}) moves along the $\Delta$-axis into the pump trajectory which, in turn, becomes topologically non-trivial.
Interestingly, the pumped charge in this case is expected to be $\Delta Q = 1$, compared to the usual $\Delta Q = 2$ in the `traditional' trajectory centred at the origin.

We develop an intuition for the interaction-induced pump by adiabatically following the ground states in the interacting Rice-Mele model.
Let us start on the left of the $(\delta,\Delta)$--plane [Fig.~\ref{fig:1}(f)] which is characterised by 
predominant singlet-correlations along the strong bonds.
Following a hypothetical singlet-pair (marked as (i) and highlighted in orange) out of the full many-body ground state, the two atoms are converted into a double occupancy for large sublattice offsets (ii, $\Delta > U/2$).
Subsequently, the double occupancy is converted back to a singlet ($\delta > 0$), albeit shifted by half a unit cell (iii).
In the last section of the pump trajectory, the pair of atoms remains in place, leading to an overall shift of half a unit cell per pump cycle (pumped charge $\Delta Q= 1$ per unit cell) while the local spin and charge correlations adiabatically adjust, ensuring return to the initial state (iv -- v).
In comparison, a non-interacting atom would only oscillate to and fro during this process (i -- v), leading to zero net current.
Note that the sketch describes an idealised adiabatic situation in which the system  remains in its instantaneous ground state at all times.
Specific aspects of the experiment and of the concrete model can affect the adiabaticity, which we will address in Sec.~\ref{sec:corrs} below.

Crucially, the lattice potential [schematics in Fig.~\ref{fig:1}(f)] does not exhibit a sliding motion, which is used in the usual Rice-Mele pump~\cite{wang_topological_2013,lohse_thouless_2016,nakajima_topological_2016}.
Instead, a `long' and a `short' lattice are slowly oscillating with respect to one another.
Consequently, this interaction-induced pump does not have a classical, non-interacting counterpart.

\section{Experimental realisation}

We use fermionic $^{40}$K atoms in a dynamically controlled optical lattice to realise the Hamiltonian in Eq.~\eqref{eqn:RM}.
Here, atoms take the roles of pumped charges.
The value of average tunnelling is $t/h = \SI{368(25)}{Hz}$, where $h$ is Planck's constant; the Hubbard $U$ is widely tuneable via a magnetic Feshbach resonance.
The lattice laser setup, derived from a single laser source at $\lambda = \SI{1064}{nm}$, is described in detail in 
Appendix~\ref{app:exp}.
In short, a combination of standing waves in $x$--, $y$--, and $z$--directions and a running wave component along the pumping ($x$--)direction superimpose to form effectively one-dimensional tubes of superlattices (size of one unit cell $\equiv d = \lambda$).
The relative phase $\varphi$ between interfering beams along $x$ and $z$, as well as the lattice depth $V_{\text{Xint}}$ of the interfering $x$-lattice give independent control over $\delta$ and $\Delta$ (see App.~\ref{app:exp}).
Prior to pumping, we maximise the proportion of doubly occupied unit cells, as described in Appendix~\ref{app:exp}, and calibrate this value to be 0.574(5), independent of the value of Hubbard $U$, where the number in brackets denotes the standard deviation.
Subsequently, pumping is initiated by sinusoidally oscillating $\varphi$ and $V_{\text{Xint}}$ out of phase with respect to one another, causing the `long' lattice (lattice spacing $= d$) to periodically move back and forth over the `short' lattice (lattice spacing $=d/2$).
The trajectory starts at $\Delta = 0$ and $\delta < 0$, then crosses the vertical axis at the \mbox{maximal $\Delta$} above the singularity, passes below the singularity, and finally returns to its initial position, as shown in Fig.~\ref{fig:2}(a).
In order to invert the pumping direction, denoted by negative time $\tau$ in Figs.~\ref{fig:2}, we start on the opposite side of the vertical axis at $\delta > 0$ and again move upwards to larger values of $\Delta$.
The `boomerang' shape of the experimental pump trajectory is a consequence of having only two modulation parameters (similar to Ref.~\cite{nakajima_topological_2016}).
In addition to varying $\delta$ and $\Delta$, the change in lattice potential leads to a variation in average \mbox{tunnelling $t$} by as much as $60\%$, but this variation does not affect any conclusions, based on our observations.

In a first experiment, we measure the in-situ centre-of-mass (c.~o.~m.) position of the atomic cloud in units of unit cells ($d$) as function of time $\tau$.
In order to quantify the transferred charge per cycle for doubly-occupied unit cells, the centre-of-mass displacement is divided by the independently calibrated filling fraction 0.574(5) described above.
The pump period is fixed to $T = 23\,\hbar/t = \SI{10}{ms}$, chosen to be much longer than the maximal inverse single-particle band gap $1/(\SI{1.4}{kHz}) = \SI{0.7}{ms}$.
In the non-interacting limit, we find no significant displacement and a linear fit yields a slope of $0.08(8)\, d/T$.
This reflects the topologically trivial nature of the pump trajectory for $U = 0$ [Fig.~\ref{fig:2}(b)].
The situation changes when performing the same experiment at a Hubbard interaction of $U/t = 9.2(7)$ [Fig.~\ref{fig:2}(c)].
Here, we measure a slope of $0.50(8)\, d/T$, consistent with the expected value of $\Delta Q = 1$ pumped charge per cycle and unit cell in a quantised, interaction-induced Thouless pump.
Compared to a usual `Rice-Mele' pump with $\Delta Q = 2$ and a measured slope of $1\, d/T$, as observed in previous experiments~\cite{lohse_thouless_2016,nakajima_topological_2016,walter_quantization_2023} and plotted as a dashed line, the interaction-induced pump transfers only half the amount of charge.
A third experiment, this time at $U/t = 16.3(1.1)$ yields no significant displacement [slope $= -0.05(8)\, d/T$], since the singularity has moved out of the pump trajectory [Fig.~\ref{fig:2}(d)]. These findings constitute the main qualitative result of our experiments.

\section{Stability of the pump}

\begin{figure}[b!]
    \includegraphics[width=0.48\textwidth]{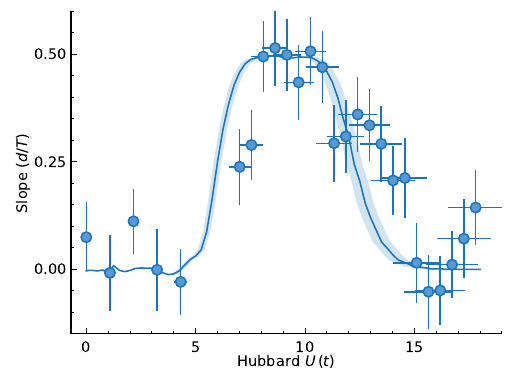}
        \caption{\textbf{Scan of interaction-induced charge pumping as a function of Hubbard $U$}. Intermediate interactions lead to an extended region of interaction-induced pumping with slopes around $0.5\,d/T$.
        The Thouless pump is rendered trivial for very strong and very weak interactions, in agreement with theory.
        Data points denote fitted slopes to time traces such as those shown in Figs.~\ref{fig:2}(b)-(d).
        Vertical error bars correspond to the standard error of the fitted slopes, whereas horizontal error bars describe the statistical experimental uncertainty in $U/t$.
        The `gap' in the data points around $U/t \simeq 6$ is due to the use of different hyperfine mixtures of $^{40}$K (App.~\ref{app:exp}).
        The solid line is a MPS calculation taking into account the experimental trajectory, as well as the trapping potential.
        The shaded area around the MPS calculation accounts for the $7\%$ relative statistical uncertainty in the experimental values of $\Delta_{\text{max}}$ and $\Delta_{\text{min}}$ (App.~\ref{app:MPS}).}
    \label{fig:3}
\end{figure}

A key aspect of our work is the ability to tune interactions and other external parameters in a controlled fashion.
In a second set of experiments we map out the parameter regions in which interaction-induced pumping occurs.
Here, we vary the Hubbard $U$ from $0$ to $18\,t$, fit lines to the data such as Fig.~\ref{fig:2}, and plot the resulting slopes in Fig.~\ref{fig:3}.
From this data we can identify three distinct regimes.
Firstly, we find vanishing displacements for interaction strengths up to $U/t \lesssim 5$ which matches the topologically trivial `pump' with zero transferred charge.
Secondly, we observe displacements of roughly $0.5\,d/T$ for intermediate interactions ($6\lesssim U/t\lesssim 11$).
The six data points on the plateau average to $0.49(3)\,d/T$ (mean and standard deviation), reflecting a large range of Hubbard $U$ for which interaction-induced pumping occurs,  consistent with the quantised value of $\Delta Q \simeq 1$.
Thirdly, a re-entrant phase appears for $U/t\gtrsim 14$ when the interactions dominate and the singularity exits the pump trajectory for strong interactions.
This interaction-induced transition to a topologically trivial situation with $\Delta Q = 0$ (or even into a nonadiabatic regime) also occurs for other types of pumps~\cite{walter_quantization_2023}, as well as interacting topological insulators~\cite{pesin_mott_2010,budich_fluctuation-driven_2013,vanhala_topological_2016}.
The observed transitions at $U/t \simeq 6$ and $U/t\simeq 13$ roughly coincide with the extremal $\Delta$--values within the pump trajectory, $2\Delta_{\text{min}}/t = 5.6$ and $2\Delta_{\text{max}}/t = 12.8$.
The scatter in the experimental data is due to drifts and shot-to-shot fluctuations of the atomic cloud, as well as our finite imaging resolution.
We measure sub-$\SI{}{\mu m}$ movements on an atomic cloud using an imaging system with a resolution of only $\SI{5}{\mu m}$ while the total cloud diameter is roughly $\SI{40}{\mu m}$.

\begin{figure}[t!]
    \includegraphics[width=0.48\textwidth]{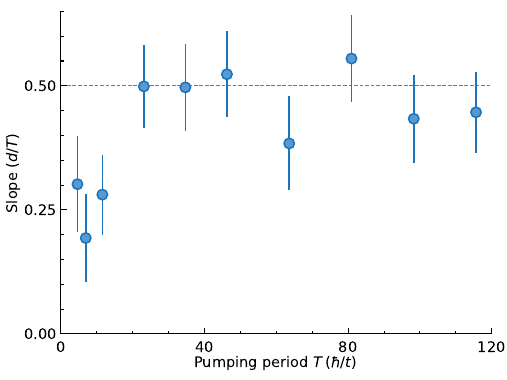}
    \caption{\textbf{Stability of interaction-induced pumping for different pump periods $T$.} For pump periods longer than 20 tunnelling times, the transferred charge per cycle becomes largely independent of $T$, supporting the existence of stable interaction-induced pumping. For shorter pump periods, the measured slopes are reduced due to non-adiabatic effects.
    Data points and error bars are analogous to Fig.~\ref{fig:3}.}
    \label{fig:4}
\end{figure}

We compare the experimental data with numerical simulations of a single, one-dimensional tube with 24 particles using MPS in Fig.~\ref{fig:3}.
These theory calculations were performed with the experimental parameters of the time-dependent Hamiltonian [Eq.~\eqref{eqn:RM}] including the trapping potential. 
The starting state was approximated by disjoint singlet states on each dimer with no long-range correlations, in order to take our loading protocol and finite temperature into account (see App.~\ref{app:MPS} for details).
The results of the $U$-scan are plotted as a line in Fig.~\ref{fig:3} and generally agree with experiment, in particular, regarding the width of the plateau of interaction-induced pumping as function of $U/t$.
The measured full-width-half-maximum of the plateau is $7.0(5)\,U/t$ in the experiment, in agreement with the theoretical value $7.2(4)\, U/t$ (value in brackets denote systematic uncertainties).
While the theory curve is slightly shifted to smaller values of $U/t$, compared with the experimental data, this is not unexpected, as the theory only takes into account a single tube and the experiment averages over a whole array of tubes with different fillings. 
Smoothening of the topological transitions is present both in theory and experiment, which is a result of nonadiabatic effects.
The presence of an extended region with displacements of close to $0.5\, d/T$ in both theory and experiment suggests that finite-entropy effects are not crucially relevant within the first pump cycle (see also Sec.~\ref{sec:corrs}).
As discussed in App.~\ref{sec:app_trap} and in Ref.~\cite{zhu_reversal_2024}, harmonic trapping affects the pumped charge only at much later times.

An important control parameter for adiabatic pumping is the duration of one pump cycle $T$, which is investigated experimentally in the following.
Keeping the interaction strength [$U/t = 9.2(7)$] and all other experimental parameters fixed, we vary the pump period $T$ over two orders of magnitude from 4 to 120 tunnelling times ($\hbar/t$).
The results are plotted in Fig.~\ref{fig:4} and the data suggests that for slow-enough pump periods the measured slope converges towards the quantised value of $0.5\, d/T$ (dashed line) and becomes largely independent of $T$.
The seven data points above $T=20 \,\hbar/t$ average to $0.48(5)\, d/T$ (mean and standard deviation).
This observation supports the conclusion that robust interaction-induced pumping occurs for a large range of parameters.
For fast pump periods below 20 tunnelling times, the interaction-induced displacement clearly breaks down due to nonadiabatic effects.

\section{Dynamics of spin- and charge-correlations during pumping}\label{sec:corrs}

So far, both the experimental data and the MPS simulations demonstrate quantised pumping in the first pump cycle for a range of finite interactions strengths, neighboured by regions of no pumping at small and large values of $U/t$, respectively.
Since there are spin-gapless excitations in the region of a nonzero pumped charge, in principle, adiabaticity is not guaranteed. Therefore, ultimately, excitations will be generated that will heat up the system, preventing quantized pumping.
We now utilise numerical simulations to explain, on the one hand, why the pumping is initially still robust, consistent with the experimental data, and 
on the other hand, to develop a microscopic picture of the nonadiabaticity in the pumping process.

Useful quantities to capture both the nature of excitations and to access heating effects are the charge and spin correlations [defined by Eqs. (A5) and (A6)].
Deviations of these correlators from the ground-state values indicate the presence of excitations and thus a finite amount of excitation energy in the system.
In a two-component Fermi gas, charge and spin are the fundamental
degrees of freedom. In some one-dimensional systems, they are fully decoupled due to spin-charge separation~\cite{giamarchi_quantum_2004}, yet, this is not the case in general. 

\begin{figure}[t]
         \includegraphics[width=0.48\textwidth]{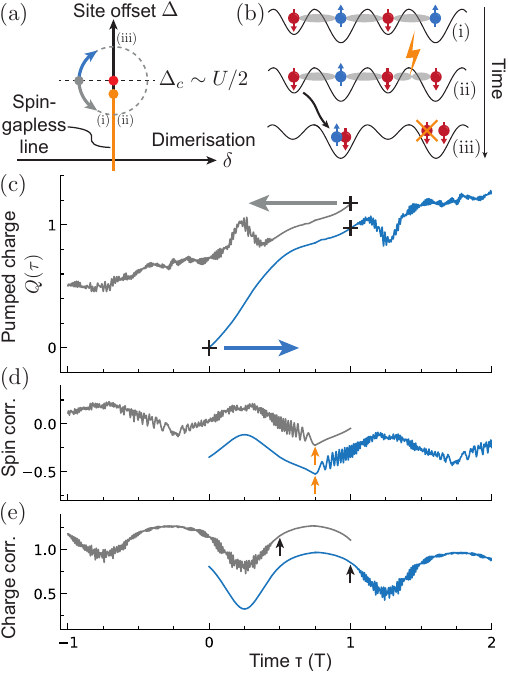}
         \caption{\textbf{Dynamics of pumped charge, charge-, and spin-correlations  computed with time-dependent Lanczos.} (a) Schematic of pump trajectory. The charge- and spin-gap closings along the ionic Hubbard axis ($\delta = 0$) are shown in red and orange, respectively. The spin-gapless line extends from $+\Delta_s$ to $-\Delta_s$. (b) Flipping the spin of the rightmost atom (a simplified example of a spin excitation) can prevent further pumping after one period. The points in time are marked with (i), (ii), and (iii) in (a) and (b). The sketch in (b) should be viewed as a conceptualised illustration and does not necessarily represent a physical low-energy excitation. (c,d,e) Dynamics of pumped charge, spin correlator, and charge correlator, respectively, calculated with Lanczos for $L=10$, $T=100\, \hbar/t$, $\delta \tau =0.1\, \hbar/t$ and antiperiodic boundary conditions (see App.~\ref{app:MPS}). The blue line corresponds to counter-clockwise pumping [blue arrows in (a) and (c)], whereas the grey lines corresponds to clockwise pumping (grey arrows, plotted in reverse time $-\tau$ for comparison). The crosses in (c) indicate quantised displacement. The orange arrows in (d) denote the crossing of the spin-gapless line. The black arrows in (e) denote the crossing of the $\Delta = \Delta_c$ line. All three grey curves have been offset on the vertical axis for clarity.}
        \label{fig:5}
\end{figure}

In our case, the ionic Hubbard model ($\delta = 0$)~\cite{fabrizio_band_1999,torio_phase_2001,manmana_quantum_2004,torio_quantum_2006} is of particular interest due to the presence of two subsequent gap closings in its many-body spectrum.
For concreteness, let us fix the value of $U/t$ and vary $\Delta$.
A gap related to reordering of the charges vanishes only 
at the single critical points $\Delta = \pm \Delta_c$ which determine the 
topology of the charge sector $[\Delta_c \sim U/2]$.
The spin gap vanishes 
for $-\Delta_s \leq \Delta \leq  \Delta_s$, where $\Delta_c - \Delta_s \lesssim t$~\cite{fabrizio_band_1999,torio_phase_2001,manmana_quantum_2004,torio_quantum_2006}, and we will call it `spin-gapless line' in the following, denoted by an orange line in Fig.~\ref{fig:5}(a).
The parameter trajectory for the interaction-induced pump crosses this line and spin excitations can occur, in principle, at zero energy cost.

The schematics in Fig.~\ref{fig:5}(b) illustrate how spin- and charge-dynamics can become intertwined during the course of a pump cycle.
While not the exact state prepared in the experiment, it is instructive to consider an anti-ferromagnetically ordered state at first [see case (i) Fig.~\ref{fig:5}(a) and (b)] and incur a spin excitation when crossing the spin-gapless line for $\Delta<\Delta_s$ (ii).
The spin excitation can take the form of a spin triplet or more complex patterns; here a spin-flip from $\uparrow$ to $\downarrow$ is shown for simplicity.
In the subsequent half-cycle, two neighbouring spins have to form a double occupancy ($\Delta>\Delta_c$) in order to ensure pumping.
However, the spin-excited pair precludes the formation of a double occupancy, thereby preventing pumping (iii).

In the following we investigate the mechanism described above using numerical simulations of pumped charge per unit cell $Q(\tau)$, spin-correlations, as well as charge-correlations as functions of time, shown in Figs.~\ref{fig:5}(c)-(e).
Here, we choose an idealised system consisting of ten lattice sites, antiperiodic boundary conditions, and an elliptical pump trajectory in order to draw generic, qualitative conclusions, leaving out some of the specifics of our experimental system (see App.~\ref{app:MPS}).
We compare two distinct trajectories, both starting on the left of the ($\delta,\Delta$)-plane, as shown in Fig.~\ref{fig:5}(a).
The `blue' trajectory crosses the spin-gapless line at $\tau = 3T/4$ while the `grey' one crosses the spin-gapless line at $\tau = T/4$.
In Figs.~\ref{fig:5}(c)-(e), the blue trajectory is always plotted forwards in time (starting from $\tau = 0$) while the grey trajectory is plotted backwards in time (starting from $\tau/T = 1$ and offset on the vertical axis) in  order to simplify the comparison.
The blue trajectory clearly shows a quantised response (black crosses) for one period, while the pumped charge is visibly reduced during the second period.
The grey trajectory follows the blue trajectory for half a period but departs afterwards.

In order to understand the charge dynamics described previously we now consider the dynamics of spin- and charge-correlations in Figs.~\ref{fig:5}(d) and (e).
The spin-correlations visibly start to oscillate when crossing the spin-gapless line after $3/4$ (blue) and $1/4$ (grey) of the pump period (orange arrows).
Indeed, the charge dynamics remains smooth up until $\Delta \sim \Delta_c$ which happens after $1$ (blue) and $1/2$ (grey) period, respectively (black arrows).
We can conclude that the spin excitations are converted to charge excitations with a delay of a quarter period.
Interestingly, this effect is largely independent of system size, as shown in Fig.~\ref{fig:si2}.
To summarise, crossing the spin-gapless line does not necessarily lead to an immediate breakdown of pumping, but excitations first have to spread to the charge sector in order to influence the pumped charge.

The spin-gapless line plays an important role in determining the long-time behaviour of the interaction-induced charge pump \cite{nakagawa_breakdown_2018}.
In the experiment, we observe a reduction of the pumped charge after the first cycle (Fig.~\ref{fig:si1}), which is indicative of a breakdown of pumping after one cycle.
Indeed, the interplay between spin- and charge-degrees of freedom implies that the dynamics of the second and subsequent pump cycles will become increasingly dependent on the precise system parameters, such as system size and pump trajectory.
We investigate these effects using numerical calculations in Fig.~\ref{fig:si3}.
The precise values of the excitation gaps matter crucially, which can be tuned by modifying the pump cycle and the driving protocol.
Altering the pump trajectory in the experiment corresponds to including higher harmonics in the driving waveform which we plan to investigate in the future.
Similarly, the addition of an Ising-type interaction term leads to an explicit opening of the spin gap in the Mott-insulating regime of the ionic Hubbard model, which can stabilise the pump over many cycles (Fig.~\ref{fig:si4}).
Ising-anisotropies have been realised with two-component bosons~\cite{jepsen_spin_2020}.

An alternative approach involves the intermediate (third)
phase in the ionic Hubbard model, called the spontaneously dimerised insulator (SDI).
The three phases of the ionic Hubbard model for repulsive interactions are, in ascending order of $\Delta$, Mott insulator, SDI, and band insulator.
The transition from a Mott insulator to the SDI is accompanied by a spin-gap opening.
The transition from the SDI to the band insulator is due to a crossing 
of two ground states with different charge distributions.
Thus, adiabatic pumping could potentially be stabilised by enlarging the SDI phase and avoiding the Mott insulator altogether.
It has been suggested that including a density-dependent hopping term into the ionic Hubbard model can enlarge the SDI phase~\cite{roura-bas_phase_2023,segura_charge_2023}.
A density-dependent hopping can be engineered by near-resonant Floquet modulation~\cite{ma_photon-assisted_2011,meinert_floquet_2016,gorg_enhancement_2018,messer_floquet_2018}.

The considerations of this section explain why adiabatic pumping is expected to break down after one pump cycle, the reason being the coupling of the spin- and charge-sectors.
Two directions for future experiments emerge: First, quantised pumping could be realised with a modified Hamiltonian (e.g., without SU(2) symmetry) and second, the present setup is ideally suited so study the interplay of charge and spin excitations in strongly correlated systems.

\section{Outlook}
In conclusion, we have experimentally demonstrated an interaction-induced charge pump using ultracold fermions in a dynamical superlattice.
The observed transport is consistent with quantised pumping within a range of repulsive interactions, while it has no classical counterpart and the pumped charge is zero below a critical Hubbard $U$.
Pumping of one charge per cycle is not possible for $U = 0$, therefore, the pumping protocol is not adiabatically connected to the non-interacting limit.
Our numerical simulations unveil the  mechanism for the breakdown of adiabatic pumping, in which spin excitations are transferred to the charge sector only later in the pump cycle.

These results open up multiple avenues for future research into topological many-body systems.
For example, the presence of trapping potentials in the experiment could lead to interaction-induced edge physics~\cite{zhu_reversal_2024}, possibly giving rise to novel topological boundary modes~\cite{rachel_interacting_2018}.
Novel cooling mechanisms in optical lattices may be enabled via density redistribution~\cite{chiu_quantum_2018}, making use of the inherent backaction between density and pump-induced currents in the interaction-induced charge pump.
In addition, the coupling mechanism between spin and charge degrees of freedom could be harnessed to realise singlet pumping, in view to realise measurement-based quantum computation~\cite{das_controlled_2006}.
In general, our work establishes Thouless charge pumping as a promising system to investigate interaction-driven physics in topological systems and topological quantum matter.
Having cross-validated theory and experiment in the limit of one-dimensional dynamics, the experimental platform can be extended to two~\cite{fu_two-dimensional_2022, padhan_interacting_2024} and even three dimensions, eventually addressing questions beyond the reach of numerical simulations.

Source data is partially available at ref.~\cite{viebahn_dataset_2024}.

\section*{Acknowledgements}

We would like to thank Kaden Hazzard and Philipp Preiss for comments on a previous version of the  manuscript.
K.V., A.-S.W., Z.Z., M.G., and T.E.~acknowledge funding by the Swiss National Science Foundation (Grants No.~182650, 212168, NCCR-QSIT, as well as advanced grant AdiaPump 209376) and European Research Council advanced grant TransQ (Grant No.~742579).
E.B. and F.H.-M. acknowledge funding by the Deutsche Forschungsgemeinschaft (DFG, German Research Foundation) – 277974659, 436382789, 493420525 via DFG Research Unit FOR 2414  and large-equipment grants (GOEGrid cluster). 
A.A.A.~acknowledges financial support provided by PICT 2020A 03661 and PICT 2018-01546 of the Agencia I+D+i, Argentina.
This research was supported in part by the National Science Foundation under Grant No.~NSF PHY-1748958.



%

\appendix

\clearpage
\newpage

\setcounter{figure}{0} 
\setcounter{equation}{0} 

\renewcommand\theequation{A\arabic{equation}} 
\renewcommand\thefigure{A\arabic{figure}} 

\section{Experimental details}\label{app:exp}

\begin{figure*}[t!]
    \includegraphics[width=0.8\textwidth]{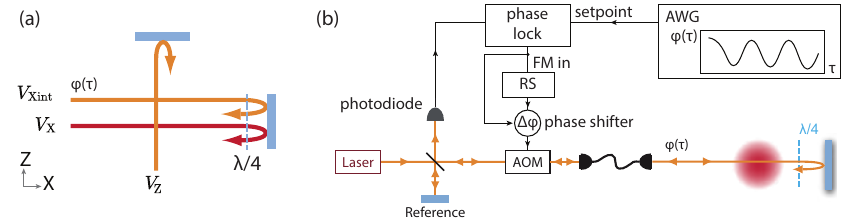}
    \caption{\textbf{Experimental setup and phase lock.} (a) Sketch of the optical lattice setup. The $y$-direction is omitted for clarity. The laser beam along $x$ contains two distinct light frequencies. $V_{\text{Xint}}$ and $V_{\text{Z}}$ (orange) describe an interference pattern in the $xz$-plane, which is sensitive to the superlattice phase $\varphi(\tau)$. $V_{\text{X}}$ is detuned with respect to $V_{\text{Xint}}$ by \SI{384}{MHz}, leading to a positional offset of half a lattice site at the position of the atoms, resulting from a $\SI{20}{cm}$ path difference between the atoms and the retro-mirror. (b) Phase lock setup. The incoming phases of both $V_{\text{Xint}}$ and $V_{\text{Z}}$ are stabilised using a Michelson-like interferometer (only one interferometer is shown here, for clarity).}
    \label{fig:si_setup}
\end{figure*}

\textit{Optical lattice setup.} The optical lattice setup is sketched in Fig.~\ref{fig:si_setup}.
In short, optical standing waves are formed in all three directions ($V_{\text{X}}$, $V_{\text{Y}}$, and $V_{\text{Z}}$) via retro-reflection.
The light frequencies of the standing waves $V_{\text{X}}$, $V_{\text{Y}}$, and $V_{\text{Z}}$ are all detuned by more than \SI{50}{MHz} with respect to each other, ruling out any interference on atomic timescales.
In addition, another beam $V_{\text{Xint}}$ is superimposed along the $x$-direction, exactly co-propagating with $V_{\text{X}}$ (emitted from the same optical fibre).
The light frequencies of $V_{\text{Xint}}$ and $V_{\text{Z}}$ are exactly the same, each defined by a Rhode \& Schwarz (RS) signal generator SMC100A, phase-locked to a \SI{10}{MHz} reference clock.
Therefore, an interference pattern is formed in the $xz$-plane, which depends on the phase difference between $V_{\text{Xint}}$ and $V_{\text{Z}}$, their respective amplitudes, and polarisations.

A quarter waveplate ($\lambda/4$) is used to imbalance ($I_{\text{XZ}}$) the incoming and retro-reflected part of $V_{\text{Xint}}$, giving rise to a directional motion of the lattice potential along $x$ for a time-varying superlattice phase $\varphi(\tau)$ (see also Eq.~\eqref{eqn:potential} below).

\textit{Phase lock.} In order to stabilise the superlattice phase $\varphi(\tau)$, a Michelson-like interferometer is used [Fig.~\ref{fig:si_setup}(b)].
The superlattice phase $\varphi(\tau)$ at the position of the atoms is measured by superimposing the retro-reflected part of $V_{\text{Xint}}$ with a short reference arm on a fast photodiode.
The measured voltage at the photodiode is filtered, mixed down to DC, and fed into the frequency-modulation input of the RS, effectively forming the integral part of a $PI$-loop. The proportional part (for small and fast changes of the phase) is realised by an additional phase shifter at the output of the RS.
An arbitrary waveform generator (AWG, Keysight 33400B) controls the setpoint for the phase lock, enabling essentially arbitrary control of $\varphi(\tau)$ in time.
Contrary to our previous work where a sawtooth wave was used~\cite{walter_quantization_2023}, we use a sinuisoidal waveform in the AWG (Eqn.~\eqref{eqn:phi} below), leading to a `rocking' interference pattern which oscillates to and fro.
A similar phase lock is used for $V_{\text{Z}}$, but without the AWG and phase shifter.

\textit{Lattice parameters.} The resulting time-dependent lattice potential is given by
\begin{equation}\label{eqn:potential}
\begin{split}
     &V(x,y,z,\tau) = \\
     &\quad- V_{\text{X}} I_{\text{self}} \cos^2 (kx + \vartheta / 2) \\
     &\quad- V_{\text{Xint}}(\tau) I_{\text{self}} \cos^2(kx)\\
     &\quad- V_{\text{Y}} \cos^2(ky) \\
     &\quad- V_{\text{Z}} \cos^2(kz) \\
    &\quad- \sqrt{V_{\text{Xint}}(\tau) V_{\text{Z}}} \cos(kz) \cos[kx + \varphi(\tau)] \\
    &\quad- I_{\text{XZ}} \sqrt{V_{\text{Xint}}(\tau) V_{\text{Z}}} \cos(kz) \cos[kx - \varphi(\tau)]~,
\end{split}
\end{equation}
where $k=2 \pi / \lambda$ and $\lambda = \SI{1064}{nm}$.
The constant lattice depths $[V_{\text{X}},V_{\text{Y}},V_{\text{Z}}]$ used in this paper are given by $[8.02(7) , 20.01(3) , 17.1(2) ] E_R$, measured in units of recoil energy $E_R = h^2/2m\lambda ^2$, where $m$ the mass of the atoms.
The values in brackets denote the standard deviations of the lattice depths over 5520 individual measurements.
Contrary to our previous work~\cite{walter_quantization_2023}, the value of $V_{\text{Xint}}(\tau)$ is time-dependent:
\begin{equation}\label{eqn:vxint}
    V_{\text{Xint}}(\tau) = V_0 [1 + A\times \sin(2\pi \tau/T)]\,.
\end{equation}
The average lattice depth is $V_0 = 0.30(2) E_R$ and the amplitude is $A = 0.68(7)$.
Likewise, the superlattice phase is time-dependent:
\begin{equation}\label{eqn:phi}
    \varphi(\tau) = \pi/2 \times [1 + \cos(2\pi \tau/T)]\,.
\end{equation}
The imbalance factors are $I_{\text{self}} = 1.00(2)$ and $I_{\text{XZ}} = 0.79(2)$.

\begin{figure}[htbp]
    \includegraphics[width=0.48\textwidth]{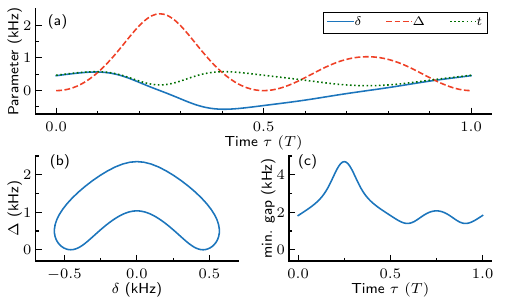}
    \caption{\textbf{Experimental pump trajectory.} (a) Rice-Mele parameters as function of time. (b) The `boomerang' pump trajectory. (c) The single-particle band gap varies during the course of a pump cycle. Its minimum is \SI{1.4}{kHz}. All energies are given in kHz ($\times h$).}
    \label{fig:si0}
\end{figure}

The time-dependent lattice parameters $V_{\text{Xint}}(\tau)$ and $\varphi(\tau)$ lead to a periodic variation of the Rice-Mele parameters $\delta$ (dimerisation), $\Delta$ (sublattice site offset), and $t$ (average tunnelling), as plotted in Fig.~\ref{fig:si0}(a).
The resulting pump trajectory is `boomerang'-shaped [Fig.~\ref{fig:si0}(b)].
In addition, we plot the minimal single-particle band gap in Fig.~\ref{fig:si0}(c).
The minimum band gap of \SI{1.4}{kHz} can be used to estimate an adiabatic timescale in the non-interacting limit.
The time-dependence of the average hopping matrix element $t$ causes a slight shift of the critical $\Delta_c$ at a fixed value of $U$ by an amount which is smaller than $t$.

A large range of Hubbard $U$ can be accessed by utilising the Feshbach resonances in the hyperfine ground-state manifold $F = 9/2$ of $^{40}$K.
For small and intermediate values of $U < 6t$, we use the mixture $m_F = \{-\nicefrac{9}{2}, -\nicefrac{7}{2}\}$.
For stronger interactions ($U>7t$), we use the mixture $m_F = \{-\nicefrac{9}{2}, -\nicefrac{5}{2}\}$.
In between the two ranges there is a small gap in the measured data points (Fig.~\ref{fig:3}).
The bracketed errors of the values of $U/t$ are dominated by the statistical uncertainty of $7\%$ in the value of $t$.

\textit{State preparation.} In order to maximise the fraction of doubly occupied unit cells we perform the following loading procedure.
First, we sympathetically cool a balanced spin mixture of $^{40}$K atoms with $^{87}$Rb, yielding $60'000(5'000)$ atoms at a temperature of $0.11(2) T/T_F$.
Second, we tune the scattering length between the spins to strongly attractive values and ramp up a deep chequerboard lattice within \SI{200}{ms}, resulting in a high double occupancy fraction.
Third, the interactions are changed to the final value and each lattice site in the chequerboard is split into two along the $x$-direction.
This loading procedure results in a filling fraction of $0.574(5)$, which is $7$ percentage points higher than in our previous work~\cite{walter_quantization_2023}.
Importantly, the population of triplet states on each double well (as well as other states such as $\uparrow\uparrow$) is negligible since  they are formed via splitting a double occupancy.

\begin{figure}[htbp]
    \includegraphics[width=0.48\textwidth]{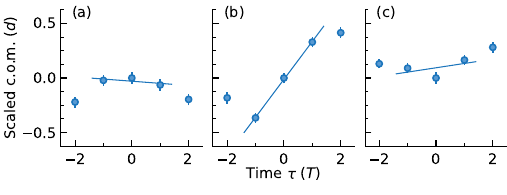}
    \caption{\textbf{Measured centre-of-mass displacements for two pump cycles.} The panels show different values of Hubbard $U/t$: (a) $0$, (b) $9.2(7)$, and (c) $16.3(1.1)$. The error bars are obtained in the same way as in Fig.~\ref{fig:2}.}
    \label{fig:si1}
\end{figure}

\textit{Observations beyond one pump cycle.} As described in the main text, we extract the scaled centre-of-mass position to measure the pumped charges.
In Fig.~\ref{fig:si1} below we plot the measured dynamics beyond one pump cycle.
The outlier in Fig.~\ref{fig:si1}(a) can be attributed to slow drifts as well as statistical noise.
Similar outliers are responsible for the scatter in Fig.~\ref{fig:3} in the main text.
Interaction-induced pumping is evident from Fig.~\ref{fig:si1}(b), but the slope does not persist beyond one pump cycle.

\begin{figure}[htbp]
    \includegraphics[width=0.48\textwidth]{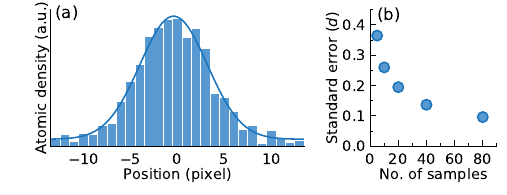}
    \caption{\textbf{Stability of centre-of-mass position.} (a) Exemplary image of the atomic cloud, integrated along the $y$ and $z$ directions. We use a two-dimensional gaussian fit to extract the centre-of-mass position. (b) When averaging over multiple realisations of the experiment, the standard error of the centre-of-mass decreases sufficiently to measure interaction-induced displacements below one lattice site.}
    \label{fig:si_stability}
\end{figure}

\textit{Stability of the centre-of-mass position.} Our experiment operates in a regime of relatively low imaging resolution (point spread function $\sim \SI{5}{\mu m}$).
Yet, we are able to discern the centre-of-mass position of the whole atomic cloud to within $ 0.1\,d = \SI{0.1}{\mu m}$.
This is possible simply by taking enough averages.
Figure~\ref{fig:si_stability}(a) shows an integrated optical density, demonstrating that the atomic cloud along $x$ (i.e.,~along the pumping direction) can be approximated by a gaussian.
The centre-of-mass position is then given by the centre of the gaussian fit [line in Fig.~\ref{fig:si_stability}(a)].
When averaging over multiple realisations of the experiment, the standard error [Fig.~\ref{fig:si_stability}(b)] approaches the value $0.1\,d$, sufficient to measure interaction-induced pumping on the sub-$\mu m$ level.

\section{Numerical simulations}\label{app:MPS}

\subsection{Real-time simulations}

For the theory curve in Fig.~\ref{fig:3}, we use a variational MPS method for ground-state calculations \cite{schollwock_density-matrix_2005,schollwock_density-matrix_2011} and time-evolving-block-decimation (TEBD) \cite{vidal_efficient_2004,white_real-time_2004,daley_time-dependent_2004} for the time-evolution.
The pumping is always started slowly via a quadratic ramp-up of the driving to minimise non-adiabatic effects \cite{privitera_nonadiabatic_2018}.
In order to make the initial state resemble experimentally realistic conditions, we start with decoupled dimer states.
Concretely, we prepare each pair of two sites (a dimer) in their ground state and then form a product of such dimers.

Technically, we start from the ground state for the decoupled Hamiltonian 
\begin{align}
   \hat H_{{\mathcal D_N}} = \sum_{(i,j)\in \mathcal D_N} \hat H_{i,j}.\label{eq:Hdimer}
\end{align}
$\mathcal D_N = \{\dots,(c-2,c-1),(c,c+1),(c+2,c+3),\dots\}$ is a set of uncoupled dimer sites centred around site $c$, such that we fill $N$ particles in total.
Each initial dimer contains 2 particles (one spin-up and one spin-down fermion). $c$ is chosen to be compatible with the dimerised ground state of the whole system.
We use an open system with size $L=49$, a trap strength of $V/t=0.034$ and $N=24$ particles. For the ground-state search, we converge the local density to a tolerance of $10^{-4}$.
The time-evolution is carried out using TEBD with a time step of $d\tau =0.01/t$ (where $t$ is the average hopping rate) and an adaptive bond dimension with a cutoff of $10^{-12}$.
The pump cycle is identical to the experimental one.
In addition, the statistical uncertainty of the experimental values of $\Delta_{\text{max}}$ and $\Delta_{\text{min}}$ is taken into account by the shaded area, which represents a scaling of the theory curve by $\pm 7\%$  relative to the solid line.
The pumped charge is computed from the centre-of-mass displacement of the cloud.

For Fig.~\ref{fig:5}, we study a finite system with open-shell boundary conditions (periodic boundary conditions for a number of sites $L$ multiple of four, antiperiodic for even $L$ not a multiple of four) to allow for the resolution of gap closings.
We consider an elliptical pump cycle 
\begin{align}
    (\Delta / t, \delta)=\left[\Delta_c + R_{\Delta} \sin (\theta), R_{\delta} \cos (\theta)\right]~,\label{eq:elliptical}
\end{align}
with $R_\Delta/t = 2.10$ and $R_\delta=0.88$.
The time-evolution is carried out with a time-dependent Lanczos method \cite{manmana_time_2005} with a tolerance of $10^{-12}$ and a time-step of $d\tau =0.01/t$. The initial state is chosen as the many-body ground state at $\theta = 0$.
The pumped charge is computed as the integral over the local particle current over one period:
\begin{align}
    Q(\tau)&=\int_{0}^{T} d \tau^{\prime}\left\langle\hat{J}\left(\tau^{\prime}\right)\right\rangle,\\
    \hat{J}&=\frac{i}{2} \sum_{j=1,2 ; \alpha=\uparrow, \downarrow}\left(t_{j} \hat{c}_{j, \alpha}^{\dagger} \hat{c}_{j+1, \alpha}-\text { H.c. }\right),
\end{align}
with $t_j(\tau)=t+(-1)^{j} \delta(\tau)$.

In order to explore the breakdown of quantised charge pumping due to low-lying spin excitations, we calculate the nearest-neighbour charge and spin correlators:
\begin{align}
  C_{\hat n}(t) &= \frac{1}{L}\langle\Psi(t)|\sum_{j=1}^L \hat n_j \hat n_{j+1}|\Psi(t)\rangle\\
  C_{\hat S}(t) &= \frac{4}{L}\langle\Psi(t)|\sum_{j=1}^L \hat S_j^z \hat S_{j+1}^z|\Psi(t)\rangle,
\end{align}
where $\hat n_j = \hat n_{\uparrow,j}+\hat n_{\downarrow,j}$ is the total particle number operator and $\hat S_j^z = (\hat n_{\uparrow,j}-\hat n_{\downarrow,j})/2$ is the total spin projection.
$|\Psi(t)\rangle$ are the time-propagated states.

In the calculations for Fig.~\ref{fig:5}, we simplify the setup and discard the trap, the variation of the hopping matrix element $t$,   and the shape of the pump cycle. As the initial state, we choose 
the ground state at the start of the cycle. We verified that this simplification does not affect the main qualitative conclusions from
Fig.~\ref{fig:5}.

We next discuss the individual effect of some of the potential  sources for nonadiabatic and nonquantised pumping. 

\subsection{Effects of trap, particle number and state preparation}\label{sec:app_trap}

As the experiment works with a harmonic trap, there are limitations concerning the maximum number of cycles that can be carried out. At the latest, when the increase of the onsite potential  due to the trap
overcomes the required variation of potentials during the pump cycle, quantised and unidirectional pumping will break down. However, this happens much later than the first pump cycle \cite{zhu_reversal_2024}.

Another possible source of imperfect pumping could be the inhomogeneous distribution of particle
numbers in the various one-dimensional systems probed simultaneously in our experiment. While it is impossible to simulate the dynamics of the full distribution for an interacting system with realistic particle numbers, we studied the behavior when increasing $N$ at fixed trap strength for the initial state
used in the experiment. All other parameters are kept fixed and correspond to the situation discussed in the main text in the context of Fig.~\ref{fig:3}. 

\begin{figure}[htbp]
    \includegraphics[width=0.48\textwidth]{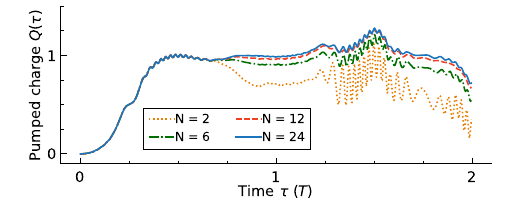}
    \caption{\textbf{Pumping for different particle numbers.} TEBD calculation for a harmonically trapped system with initial strings of dimer states with different particle numbers $N=2,6,12$, and $24$. The pump charge is calculated via the centre-of-mass displacement. Calculated for the experimental pump cycle with $L=49$, $U/t=8$, $T=23.3\hbar/t$, $d\tau=0.1\hbar/t$ and $V_0/t=0.034$.}
    \label{fig:si2}
\end{figure}

Our results are shown in Fig.~\ref{fig:si2}. Clearly, regardless of how many particles are placed into the trap (as long as we do not reach the region of a steep potential increase), during the first three quarters of the pump cycle, the behavior is largely independent thereof and the particle number affects the pumped charge only mildly. After crossing the gapless line, however,
the pump efficiency  depends on $N$, with smaller $N$ being detrimental to efficient pumping.
Notably, the pumped charge quickly becomes independent of particle number for $N\gtrsim 12$ and the pumped charge converges to a quantised value in the first pump cycle. Since particle numbers are much higher in most of the one-dimensional systems realised in the experiment, we conclude that variations in $N$ can be excluded as a dominant source of imperfect pumping on the time scales investigated.

Another potential source of imperfect pumping could be initial-state-preparation defects, e.g., hole and doublon defects as well as empty unit cells in the string of dimers. Exemplary simulations show that every string of dimers pumps independently on the relevant time scales and hence these defects
are not expected to play a crucial role.

\subsection{Role of the pump cycle}

For a strictly adiabatic situation, the choice of the pump cycle does not matter.
In our situation, where a spin-gapless region inevitably exists in the thermodynamic limit, a dependence on details of the
pump cycle is, in principle, expected. Moreover, the experiment operates at a finite entropy density.

For a finite-size system, there is, strictly speaking, no gapless continuum of spin excitations yet, hence the minimum finite-size gap along the pump cycle should matter.
To address this point, we compare 
the boomerang-shape pump cycle used in the experiment (including the actual variation of the average hopping matrix element) to an elliptical  cycle as used for Fig.~\ref{fig:5} where we keep the average hopping matrix element
constant, both with $N=10$ particles. These simulations are carried out for a closed system using the time-dependent Lanczos technique for a slow pumping protocol. We remove the trap to single out the effect of the minimum gap.

\begin{figure}[htbp]
    \includegraphics[width=0.48\textwidth]{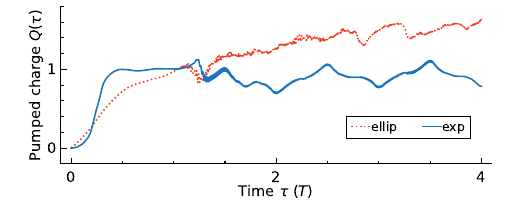}
    \caption{\textbf{Pumping with different trajectories and system sizes.} Time-dependent Lanczos calculation of the pumped charge $Q(\tau)$ in an open-shell antiperiodic system, started from the ground state. Experimentally realised pump-cycle (`exp') and elliptical pump cycle (`ellip') $L=10$. $U/t=10$, $d\tau = 0.01 \hbar/t$, $T=100\hbar/t$. 
   }
    \label{fig:si3}
\end{figure}

The results presented in Fig.~\ref{fig:si3} illustrate that during the first pump cycle, there is no
significant difference between the two pump paths. Beyond that, the behavior is strongly path-dependent, but not quantised in either case, yet somewhat larger for the elliptical cycle.
This is reflected in the minimum gaps along the two cycles, which are $\Delta E_{\mathrm{exp}} = 0.04t$, $\Delta E_{\mathrm{ellip}}=0.18t$.

One may further wonder about the effect of total particle number at fixed filling.
For the elliptical cycle 
with $N=L=6$ and $10$, the respective minimum gaps are
$\Delta E_{L=6} = 0.28t$, $\Delta E_{L=10} = 0.18t$, respectively. Consistently, the pump efficiency for $L=10$ is much worse than for $L=6$ (results not shown).

\subsection{Effect of dimer initial state instead of ground state.}

\begin{figure}[htbp]
    \includegraphics[width=0.48\textwidth]{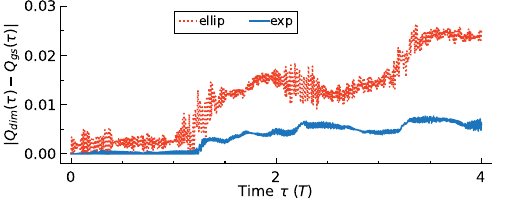}
    \caption{\textbf{Starting from the ground state versus dimer state.}
Difference of the pumped charge between a pump starting from the ground state ($Q_{gs}$) and one starting from the dimer state ($Q_{dim}$)
for the experimental and the elliptical pump cycle, $L=6$, $U/t=10$, $d\tau = 0.01\hbar/t$ and $T=100\hbar/t$.  
   }
    \label{fig:si5}
\end{figure}

In the experiment, the loading scheme leads to an initial state that can be approximated by a product of dimers, each with two particles and in their ground state. We compare pumping starting from either this state or the ground state, from the same point in the pump cycle.  The results of time-dependent Lanczos calculations are shown in Fig.~\ref{fig:si5}. The difference in the pumped charge is initially small yet increases significantly once the minimum gap is crossed. The comparison between the two different pump cycles -- experimental versus elliptical one -- shows that the former leads to smaller differences in the first pump cycle  because the starting point has a larger value of $\delta$ and therefore a stronger dimerization.

\begin{figure}[htbp]
    \includegraphics[width=0.48\textwidth]{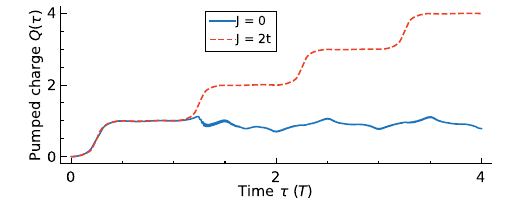}
    \caption{\textbf{Pumping with an additional interaction term.} Time-dependent Lanczos calculation of the pumped charge $Q(\tau)$ from the ground state in an open-shell antiperiodic system for the experimentally realised pump-cycle,  $L=10$, $U/t=10$, $d\tau = 0.01\hbar/t$, $T=100\hbar/t$ and two different Ising-coupling strengths ($J=0$ and $J=2t$).}
    \label{fig:si4}
\end{figure}

\subsection{Improving the robustness of the interaction-induced pump via opening the spin gap}

As already discussed in Ref.~\cite{bertok_splitting_2022}, the SU(2) symmetry of the ionic Hubbard model can be lifted by various perturbations which at the same time gap out the spin sector. This leads to robust quantised pumping over many cycles. 

We here demonstrate that this prediction remains valid also for the experimental pump cycle, by adding an Ising interaction to the Hamiltonian (other examples were studied in Ref.~\cite{bertok_splitting_2022}):
 \begin{align}
        \hat{H}_{\mathrm{Z}}=J \sum_{j=1}^{L} \hat{S}_{j}^{z} \hat{S}_{j+1}^{z}.
    \end{align}
The comparison between a simulation with and without this Ising term is shown in Fig.~\ref{fig:si4}. The results 
establish that opening the spin gap significantly stabilises the pump, leading to  robust  pumping for many pump cycles compared to the bare Rice-Mele-Hubbard model. 

In summary, quantised pumping could be achieved by lifting the spin gap, which requires experimental changes such as utilising a different atomic species~\cite{jepsen_spin_2020}.

\end{document}